# Characterisation of fire-damaged batteries – implications for recycling

Wafaa Al-Shatty,[a,b] , Tom Dunlop[a] , Rhys Charles [a] , Davide Deganello [a] , Jenny Baker*[a,b]

[a]Faculty of Engineering and Design, University of Bath, BA2 7AY, UK
[b]Faculty of Science and Engineering, Swansea University, Bay Campus, Swansea SA1 8EN, UK.

* **Corresponding author e-mail: was38@bath.ac.uk, (Wafaa Al-Shatty).**

*ORCID ID: Wafaa Al-Shatty: 0000-0002-3389-0199; Tom Dunlop: 0000-0002-5851-8713, Rhys Charles: 0000-0001-8341-4177, Davide Deganello:0000-0001-8341-4177, and Jenny Baker 0000-0003-3530-1957.

### Abstract

As lithium-ion battery demand grows, so do fire safety challenges. Despite this, research on fire-damaged batteries remains limited. This study explores the distribution of valuable metals (such as Ni, Mn, Co, Cu) in two types of waste derived from lithium-ion nickel-manganese-cobalt oxide batteries (NMC811): black mass (BM) and fire-damaged waste (FD). It emphasizes that cobalt, manganese, and nickel-rich NMC811 particles are predominantly found in smaller particle size fractions (<125 µm), where they can account for up to 85 % of total metal content. Fire-damaged (FD) batteries show a similar, though less pronounced, trend. Evidence of structural degradation suggests that fire temperatures exceeded 500°C; however, the presence of residual organic binders indicates that heat was unevenly distributed during the fire. FD batteries become friable and easily fragment into fine particles, which can hinder the effective separation of copper and aluminium current collectors, increasing their presence in processed material. The inclusion of FD batteries in standard BM processing introduces variability in output composition, potentially lowering the concentration of high-value NMC811 materials present. To maintain product quality and recycling output values, it is recommended that FD batteries are processed separately. Alternatively, particle size separation may allow for tailored outputs aligned with specific customer requirements.

### Introduction

To meet Paris Climate Agreement commitments of reducing $CO_2$ emissions to net zero by 2050, it is necessary to increase the proportion of electricity supplied by intermittent renewable sources to electrify transportation. This has led to increased demand for electrochemical energy storage. Typically, this demand is met by lithium-ion batteries (LIBs), which require ongoing supplies of ‘critical raw materials’ (CRMs), including lithium, cobalt, nickel, and graphite. Liang *et al.* calculated that some net-zero scenarios would cause a sixteenfold increase in CRM demand by 2050 from 2020 levels [1]. Therefore, meeting net zero targets will require a great increase in the efficiency with which CRMs are recycled from wastes to increase secondary supply and alleviate demand on primary CRMs [2, 3]. In the near term, recycling of LIB will predominantly target manufacturing scrap and batteries that have prematurely reached end-of-life (EoL).

The majority of electric vehicle (EV) batteries will reach EoL 15-20 years after being placed on the market [4, 5]. It is likely to be economically and environmentally beneficial to use some EV batteries in a second-life stationary application, which further delays recycling [6]. However, batteries which are removed from service early due to damage which renders them unsuitable for second-life applications will be available for recycling much sooner. In many cases, these will be fire-damaged (FD) batteries, retired prematurely following damage from fire, which may have originated in the battery pack itself, or externally, e.g. in another waste battery with which batteries are stored during waste management. EoL batteries are currently collected, discharged, and usually mechanically shredded, with plastic and metallic streams recovered and recycled directly [7, 8]. The residual battery waste takes the form of black multiphase mineral powder, called black mass (BM). Within the BM, metals account for the majority of recoverable value, particularly nickel, manganese, and cobalt. However, BM also contains a significant portion of uncharacterised residual materials [9-11]. There is no defined specification for BM, and recycling yield and purity are affected by processing parameters such as the initial shredding size [12]. FD batteries may therefore be processed within mixed batches containing non-FD batteries.

The impact of FD will depend on several factors, including whether thermal runaway occurred [13, 14], the peak temperature reached, and what fire retardants were used whilst extinguishing the fire. Thermal studies of common EV LIB chemistries such as lithium iron phosphate (LFP) and lithium nickel manganese cobalt oxide with the ratio 8 nickel, 1 managnese 1 cobalt (NMC811) have shown that the thermal response of EOL battery materials can be divided into four stages [15-20]:

i) 20-170°C: organic components, including electrolytes, volatilise.
ii) 300-475°C: decomposition of residual organics occurs with the binder (Polyvinylidene fluoride (PVDF)) decomposing at 300°C and the separator at 475°C [18].
iii) 500-800°C: [19], carbon black decomposes.
iv) 800-1000°C: decomposition of cathode active material [20].

Research on the potential of recovering materials from FD batteries is limited. In some cases, they are processed within mixed batches of undamaged batteries, but the effect on BM output quality is unknown. This article characterises FD batteries to help recyclers predict this impact. As this study focuses specifically on NMC811 chemistry containing high-value metals such as Ni and Co, the economic and material impacts of FD are particularly significant. Other chemistries without Ni or Co, such as LFP ($LiFePO_4$), exhibit different thermal behaviour during fires and therefore the findings presented here should be interpreted within the context of NMC811-based systems.

### *1.1 Methods*

BM used in this research was supplied by Treharne EV Recycling (Wales, UK), produced by processing EoL NMC811 batteries using a proprietary method involving crushing and subsequent separation of plastics and metal foil/casings. Chemical assay data supplied with the BM is given in Table 1. The BM was sieved into three powder fractions: > 300 µm, 125-300 µm, and < 125 µm. Fire-damaged NMC811 battery samples (FD) were received from Treharne EV Recycling without pre-treatment, and are known to originate from multiple NMC811 lithium-ion cylindrical cells; no further information regarding fire conditions, suppression method, or number of source cells/modules was available. The samples were manually crushed using (mortar and pestle), then the powder was sieved into fractions of the same particle sizes for further study. Both waste streams are compared to commercial NMC811 (179802-95-0) acquired from Ossila.

Quantitative analysis of metal content was conducted by MP-AES according to methods outlined by Agilent Technologies [21]. Solutions were prepared by acid digestion of 0.2 g powder samples, by boiling in 40 mL of aqua regia ($HNO_3$: HCL, 3:1) for 2 hours. Nitric acid (ASC reagent ≥ 90.0 %) and hydrochloric acid (ASC reagent, 37 %) were purchased from Sigma Aldrich. The digested solution was filtered through a 0.45 µm filter, and the resulting filtrate was passed through a fresh 0.45 µm filter a second time to ensure complete removal of undissolved solids. The resulting filtrates were diluted with deionized water (18 MΩ·cm) to a final volume of 100 mL in a volumetric flask for MP-AES analysis. ICP standards sourced from Sigma Aldrich, including silicon (08729), phosphorus (38338), and multi-element standard solution IV. (1.11355.0100) were used for MP-AES calibration. All chemicals were used as received.

Table 1: Composition of BM (supplied by Treharne)

| Materials/elements | % by weight |
|---|---|
| Nickel (Ni) | 25-35 |
| Manganese (Mn) | 3-4 |
| Cobalt (Co) | 4-5 |
| Copper (Cu) | 0.5-1 |
| Aluminium (Al) | 0.2-0.5 |
| Silicon (Si) | 1.5-2.5 |
| Iron (Fe) | 0.14-0.18 |
| Phosphorus (P) | 0.25-0.35 |
| Carbon and lithium-ion | 55-60 |

A Bruker D8 Discover with a point source (Cu Kα λ = 0.154187 nm) at 40 kV and 40 mA and a polycapillary element was used for XRD analysis. Thermal XRD was performed with an Anton Paar DHS 1100 high-temperature chamber, purged with nitrogen gas and a heating rate of 10 °C /min followed by a 10 minute hold before acquiring the XRD scan over a 2θ of 10°–70° and a step size of 0.03°. Reference data from the ICDD PDF-2+ database and supported by the Crystallography Open Database (COD) was used for phase identification. for comparison and verification.

SEM was performed using a JSM-7800F Schottky Field Emission SEM with an Oxford instrument X-Max EDS detector and a SU3900 Variable pressure SEM also equipped with an Oxford instrument X-Max EDS detector. Thermal analysis was performed using SDT Q600 Simultaneous TGA/DTA, 20 ± 2 mg of samples were heated from 25°C at a rate of 10 °C/min to 1200°C with a 100 mL/min flow rate of air or high-purity (99.99 %) argon carrier gas with anaverage of three repeats. Resonant Raman microscopy was performed using a Renishaw inVia™ Raman microscope, with a 633 nm laser and a Leica PL Fluortar L50x/0.55 long working distance objective lens between 1000 to 3000 $cm^{-1}$ Raman shift. The laser beam was focused by maximising the G-peak intensity to confirm the best z-height alignment of the beam between the sample and detector. An average of three locations was taken, and the data was analysed using WiRE5.0 software.

### *1.2 Results and Discussion*

The received samples and sieved fractions are shown in Figure 1, and the particle size distribution of the sieved fractions is given in Table 2. The jelly roll cell structure of the FD batteries was clearly visible in the received sample prior to grinding (Figure 1a.) Similar quantities of fines (<125 μm) are present in both samples, which is expected as fines are inherent in the battery anode and cathode materials, and the shredding process will not have significantly altered the proportions present. The ground FD sample, however, yields a far greater proportion of larger particles (> 300 μm) than the BM sample, which is attributed to the lack of mechanical grinding and shredding in preparation of the FD material and less effective size reduction achieved with manual grinding with pestle & mortar.

Table 2: Particle size distribution of BM and FD sieved fractions.

| Particle size (μm) | Wt.% | |
|---|---|---|
| | BM | FD |
| >300 | 13.7 | 43.3 |
| 125-300 | 50.0 | 25.8 |
| <125 | 36.3 | 31.0 |

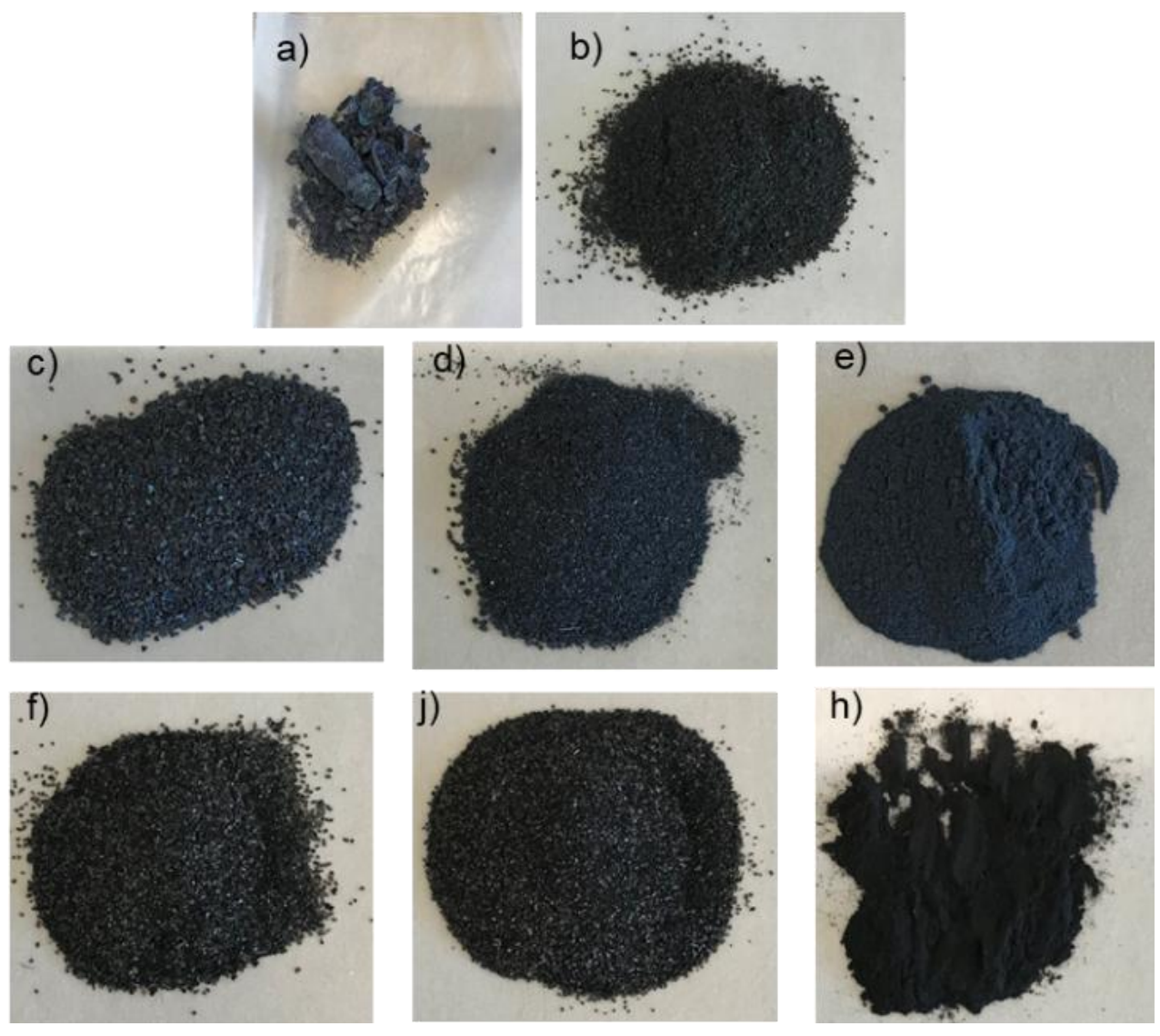

Figure 1: Photograph of analysed samples: a. as received FD sampleb. as received BM c. >300 μm FD fraction; d. 125-300 μm FD fraction; e.<125 μm FD fraction; f. >300 μm BM fraction; g.125-300 μm BM fraction; and h. <125 μm BM fraction.

Figure 2a shows the XRD spectra of the sized fractions of BM. The NMC811 structure is observed at all particle sizes, although at a lower peak intensity in the larger particle sizes. This result is mirrored by the MP-AES results (Table 3) which show approximately double the amount of Li, Co, Ni and Mn in the smaller particle sizes compared with the larger particles.

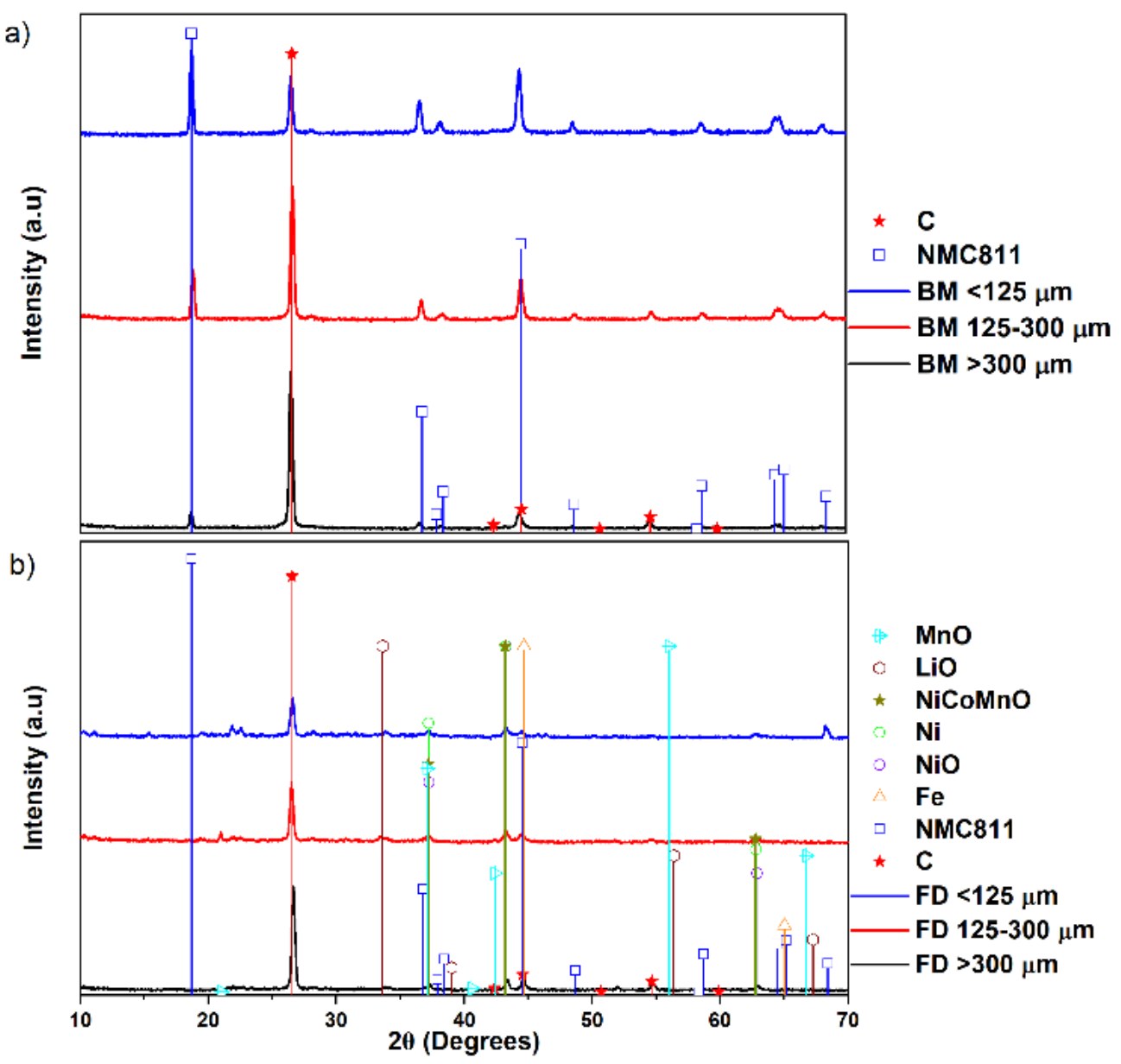

Figure 2: Comparison of XRD diffractograms for a. BM material, and b. FD material at different particle sizes.

This suggests a preferential concentration of the cathode material in the fines. Conversely, copper and aluminium originating from current collectors, which have been ineffectively separated in the mechanical separation of shredded batteries, are present in greater concentration in larger-sized fractions. This is likely due to the ductility of these foils preventing extensive size reduction during grinding and shredding, such that they will deform and ball up rather than reduce in size. MP-AES results of all size fractions of the BM align with the assay provided by the supplier of the BM.

The XRD patterns of the <125 μm FD fraction (Figure 2b) exhibit a dominant and sharp graphite (002) peak at approximately 26.5°, and indicates the preservation of the graphitic structure despite exposure to the fire. However, the characteristic peaks associated with the layered NMC811 phase—typically observed at 18.7°, 36.5°, 44.7°, and 65.5° 2θ, corresponding to the (003), (101), (104), and (108) planes—are either significantly diminished or absent. This suggests substantial structural degradation or loss of crystallinity in the NMC811 phase during the material handling or separation process. MP-AES results confirm the cathode metals are concentrated in the fines fraction; therefore, the suppression of these peaks may be attributed to particle fracture, surface amorphization, or partial phase transformation due to mechanical stress or thermal exposure, reducing long-range order within the NMC811 lattice. The absence of detectable secondary phases further implies that the NMC811 structure has been disrupted but has not converted into a distinct crystalline degradation product detectable by XRD.

MP-AES analysis shows significantly higher copper and aluminium present in the FD sample compared with BM, which is unsurprising in light of the fact FD batteries have not gone through the proprietary separation process which produced the BM by separating casings and current collectors. Elevated levels of metals from casings and current collectors in materials destined for downstream recovery via hydrometallurgy can cause process issues. Side reactions during acid leaching consumed reagents, increasing process costs and forming insoluble residues (e.g., $AlF_3$, Al $(OH)_3$). These species can reduce the selectivity of metal recovery and complicate downstream solvent extraction by introducing additional competing ions. To mitigate these effects, preprocessing steps such as sieving, density separation, eddy current and magnetic separation targeting removal of conductive foils before leaching would be beneficial [22].

The BM has spherical particles of 4-7 μm (comparable to the unprocessed commercial NMC811 powder -Figure S1, and XRD patterns Figure S2) present in all fractions (Figure 3a-c). This indicates the NMC811 within BM retained its morphology during processing (shredding, separation, and sieving, etc) [23, 24]. Figure 3d-f shows the surface morphology of the FD sample size fractions. The larger particles show damage to the spherical structure (highlighted in blue in Figure 3d), indicating that whilst in some instances the morphology is retained as the NMC811 decomposes, the larger particles exhibit surface damage. Since there is no fundamental damage to the materials structure of NMC811 and Graphite within the BM, whilst in the FD data the NMC811 exhibits damage and therefore this is attributed to the fire since the FD it has not undergone any post processing which would cause this damage.

Figures S3–S5 and Table S1 (supporting document) present the SEM-EDS elemental analysis of FD samples categorized by size (>300 μm, 125-300 μm, and <125 μm) prior to XRD thermal treatment. The accompanying map illustrates the local spatial distribution of elements such as Ni, Mn, and Co, superimposed on the relevant SEM images. The maps show that the coarse fraction (>300 μm) contains localized Ni-rich agglomerates, whereas the finer fractions (125-300 μm and <125 μm) are dominated by Cu- and Al-rich regions, consistent with current collector fragments.

Table 3: Elemental analysis of BM and FD size fractions measured by MP-AES after acid digestion*. Results are an average of two samples ± 1 standard deviation

| | **Particle size (μm)** | **>300** | | **125-300** | | **<125** | |
|---|---|---|---|---|---|---|---|
| **Element** | **Sample** | **BM** | **FD** | **BM** | **FD** | **BM** | **FD** |
| **Li** | | 4.800 ± 0.100 | 1.500 ± 0.900 | 6.900 ± 0.700 | 2.3000 ± 0.300 | 9.900 ± 0.600 | 2.000 ± 0.400 |
| **Al** | | 1.200 ± 0.200 | 1.300 ± 0.800 | 0.750± 0.030 | 1.023 ± 0.050 | 0.560 ± 0.080 | 1.400 ± 0.100 |
| **Cu** | | 4.190 ± 0.080 | 8.000 ± 5.000 | 2.214 ± 0.005 | 9.200 ± 0.300 | 1.770± 0.090 | 16.000 ± 2.000 |
| **Co** | | 3.530 ± 0.060 | 1.300± 0.600 | 5.100 ± 0.400 | 2.000 ± 0.300 | 7.800 ± 0.500 | 1.100 ± 2.000 |
| **Fe** | | 2.400 ± 0.300 | 0.300± 0.200 | 1.150 ± 0.020 | 0.900 ± 0.300 | 2.610 ± 0.080 | 1.700 ± 0.100 |
| **Mn** | | 1.903 ± 0.003 | 0.900 ± 0.500 | 2.700 ± 0.400 | 1.114 ± 0.006 | 4.200 ± 0.300 | 0.680 ± 0.050 |
| **Ni** | | 28.900 ± 0.200 | 11.000 ± 5.000 | 40.000 ± 3.000 | 16.700 ± 0.400 | 60.000 ± 4.000 | 9.000 ± 1.000 |
| **P** | | 0.086 ± 0.001 | 0.300 ± 0.200 | 0.067 ± 0.002 | 0.390 ± 0.010 | 0.055 ± 0.002 | 0.550 ± 0.050 |
| **Si** | | 0.066 ± 0.004 | 0.040± 0.030 | 0.070 ± 0.010 | 0.061 ± 0.006 | 0.110 ± 0.050 | 0.071 ± 0.003 |

*The reported values represent **mass-averaged (global) compositions** of each size fraction and therefore include contributions from both active material and residual non-active components (for example, Al- and Cu-rich current collector fragments), particularly in the FD samples. Local elemental enrichment observed by SEM–EDS is not directly reflected in these bulk measurements.

Because the FD material was not subjected to the standard black-mass separation process, non-active components remain present and contribute to dilution effects observed in bulk compositional analyses. The enrichment of nickel in the coarse fraction (> 300 µm) of the FD sample indicates preferential partitioning of Ni into larger particles following fire exposure. One possible explanation is thermally induced segregation under locally reducing and high-temperature conditions, where partial reduction of $Ni^{3+}$to $Ni^{2+}$ or metallic Ni may facilitate sintering and coalescence of Ni-rich domains. However, this mechanism remains inferential in the absence of direct phase or valence-state characterization.

In contrast, Mn- and Co-rich phases are more thermally stable or partly volatilized, leading to their relative depletion in smaller size fractions. Consequently, the coarse particles exhibit higher nickel content, reflecting both compositional redistribution and physical segregation that occurred during the fire event. Table S2 summarizes the key factors influencing particle size and composition in FD NMC811 batteries, the underlying effects on the material, and resulting outcomes relevant to BM characteristics. Figures S6–S8 and Table S3 illustrate EDS elemental mapping of BM before XRD thermal treatment. The data indicate higher concentrations of cathode metals in all fractions compared to FD samples (Table S3), consistent with the MP-AES data.

The Raman shift of the BM fractions shows a decrease in the G peak intensity relative to the D peak, indicating a reduction in graphitic nature in all except the largest size fraction Figure 4a [25]. This suggests graphite is concentrated in the largest size fraction, and smaller fractions contain more non-graphitic carbon, such as carbon black.

The FD material has a lower D: G ratio than equivalent-size BM fractions, suggesting that there has been further degradation of graphitic structure during the fire (Figure 4b).

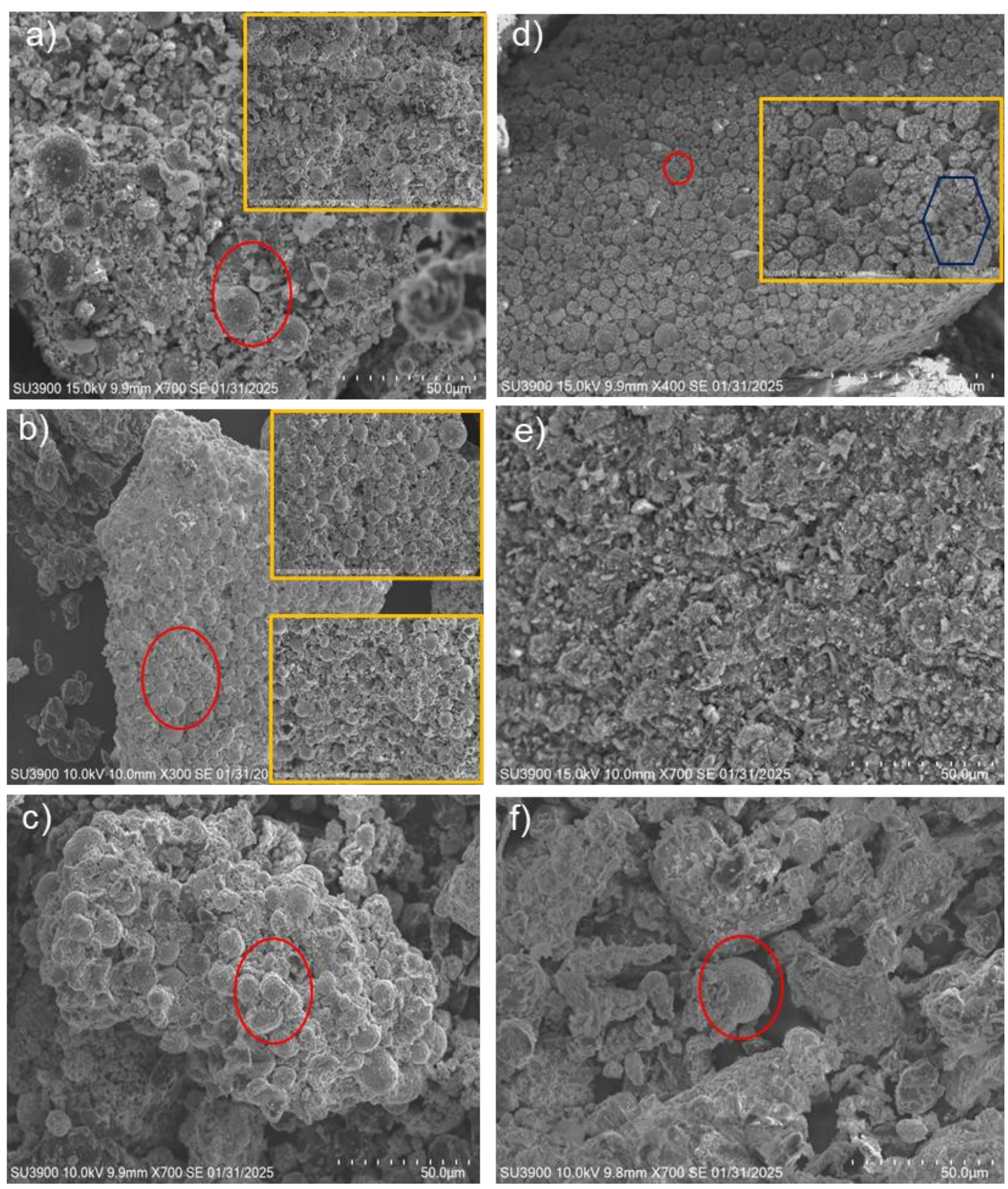

Figure 3: SEM images of BM at three fractions (a-c), (>300 µm, 125–300 µm, and <125 µm), and (d-f) of FD (>300 µm, 125–300 µm, and <125 µm). This data was acquired by the SU3900 Variable Pressure Scanning Electron Microscope.

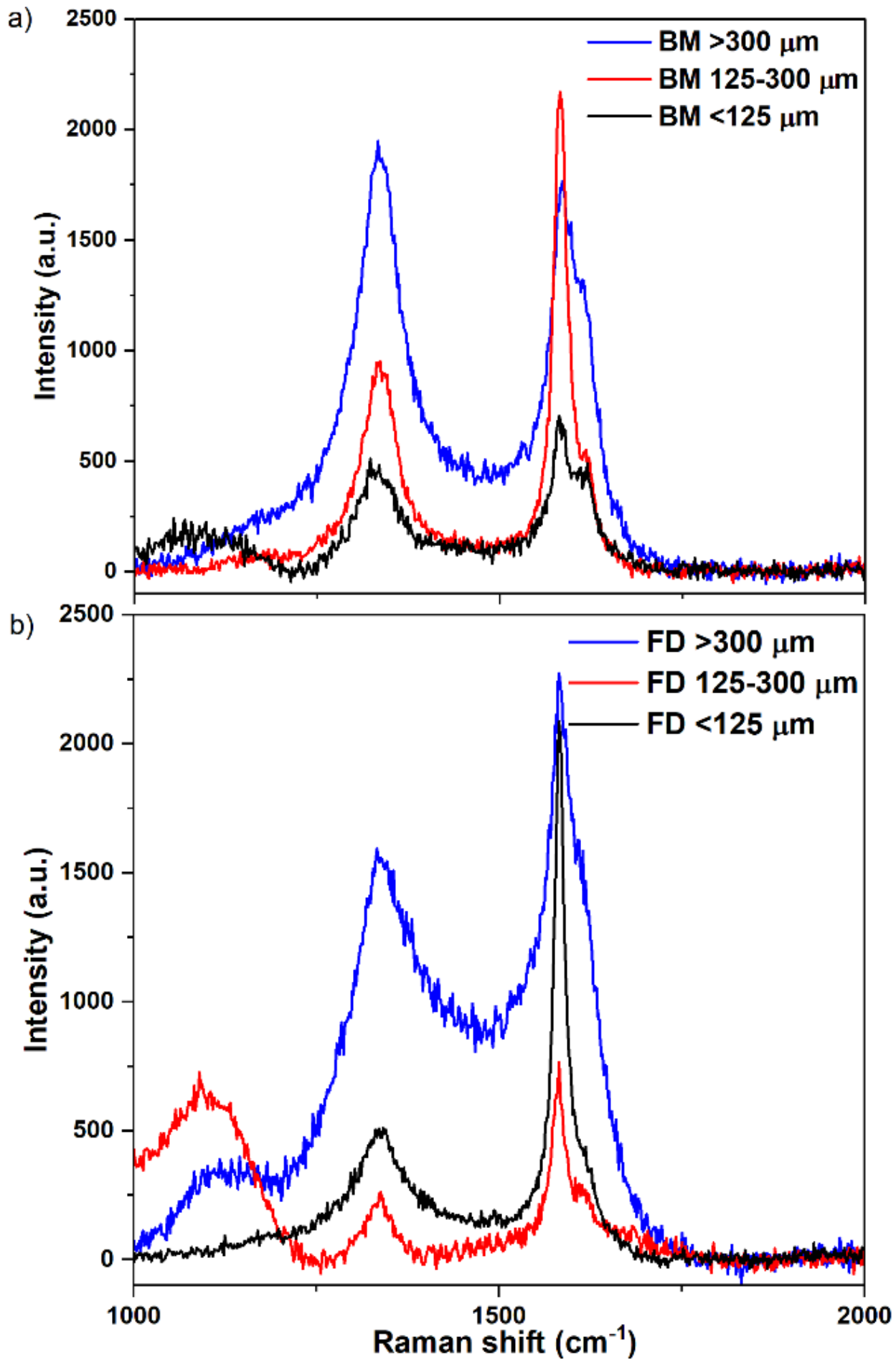

Figure 4: Raman shift of sized fractions of a. BM and b. FD material.

The TGA profile of the BM sample exhibits two weight loss steps (Figure 5a and Table 4). From 25 to 500°C, ~10 % weight loss in all size fractions is seen, resulting from decomposition of organic materials (such as binders). From 500 to 800°C, larger-sized particles have higher weight loss compared to finer particles. The mass loss at ~600°C is associated with the NMC811 transformation [15, 26]. Total mass loss in smaller size fractions is lower than in larger-sized fractions, indicative of higher NMC811 content, which decomposes to stable metal oxides. This mirrors the XRD results.

The FD profile exhibits a similar trend in the loss of organic materials, indicating that while parts of the battery likely reached sufficient temperatures to reduce NMC811 to Ni and Co metal (>500°C), the presence of residual organics suggests not all of the material reached such high temperatures. During fire exposure, residual electrolyte and binder decompose below 500°C, releasing CO, $H_2$, and HF, creating a locally reducing environment which promotes reduction of $Ni^{3+}$ to $Ni^{2+}/Ni^{0}$ and facilitates LiF formation through reaction of Li with HF. The non-uniform temperature distribution across the batteries has resulted in heterogeneous decomposition and particle size-dependent secondary phase formation in NMC811.

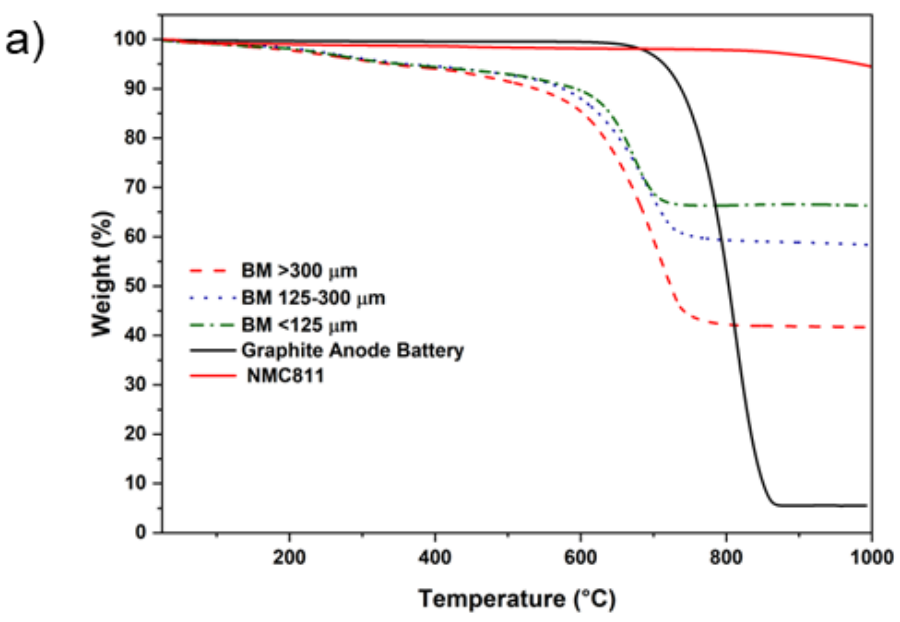

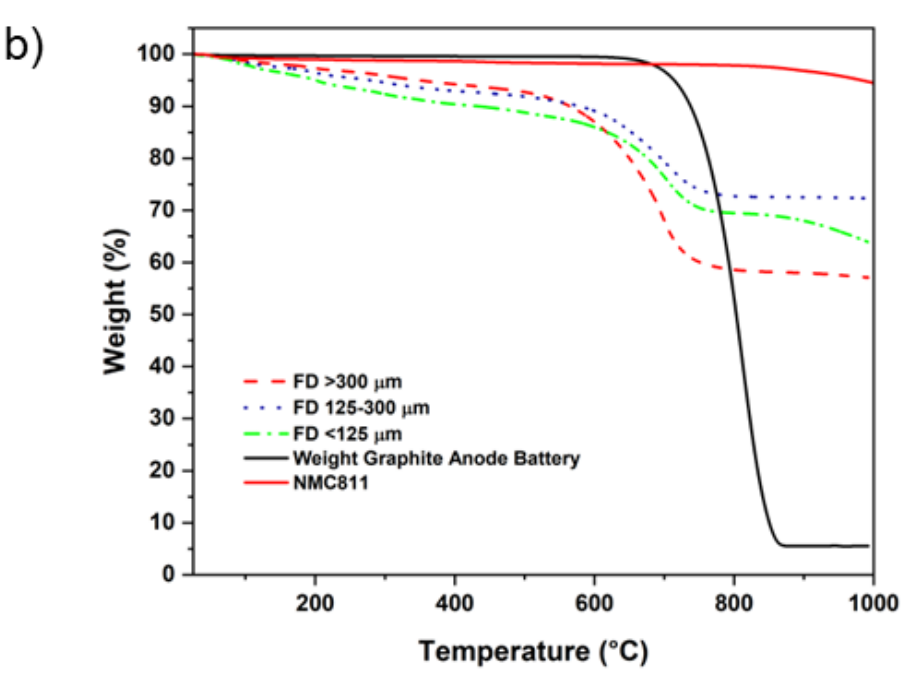

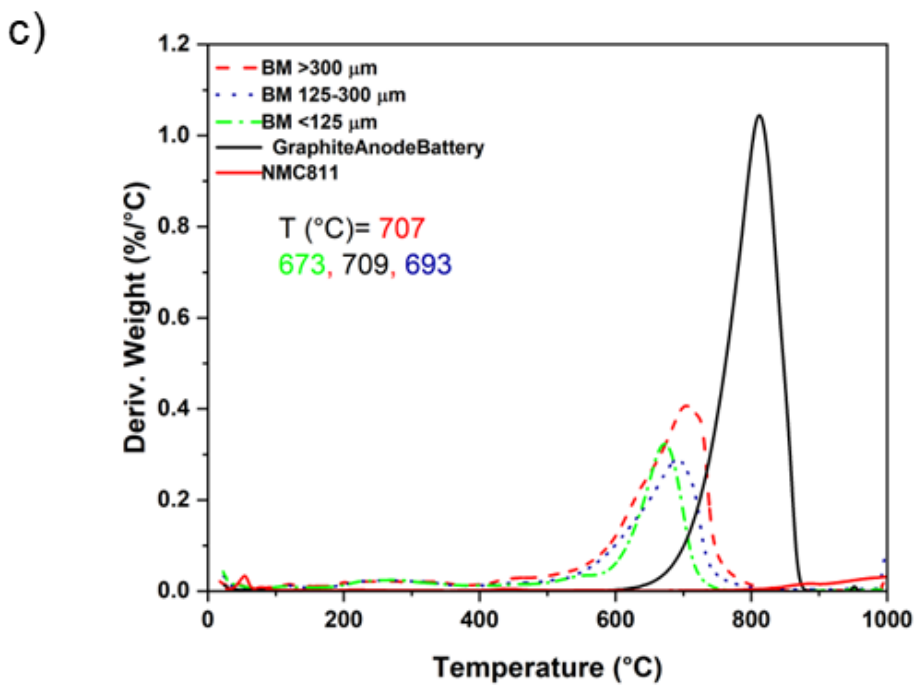

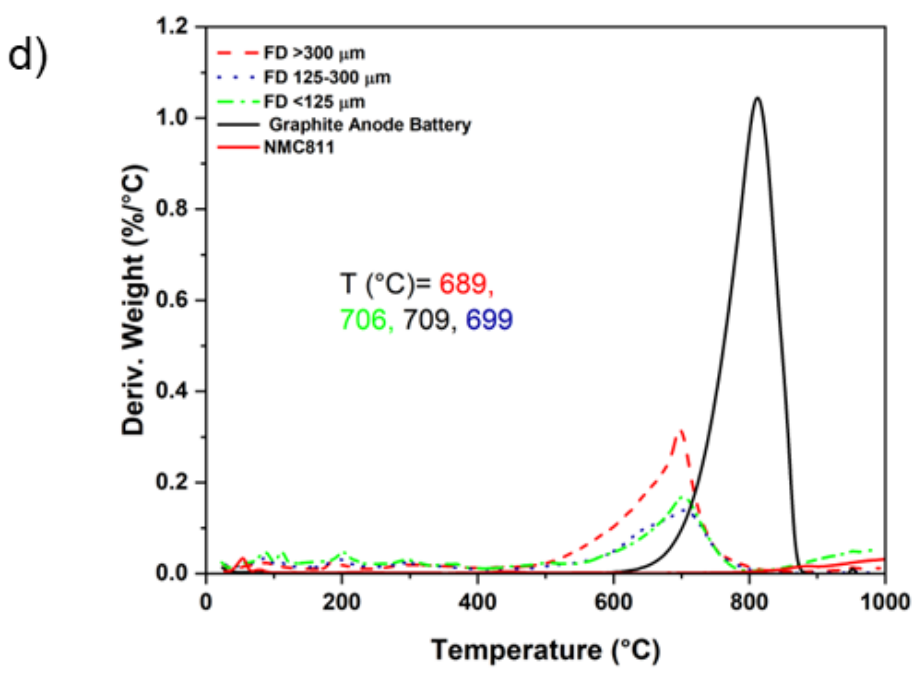

Figure 5. Thermal analysis in air: a. TGA of BM, b. TGA of FD, c. derivative of TGA BM, d. derivative of TGA FD. At 10°C/1000°C, at a flow rate of 100 ml/min. Average of the three repeats.

Table S4 highlight observed phenomena in FD batteries, their mechanistic effects, and resulting changes in BM composition and microstructure, highlighting potential challenges for downstream recycling processes. Figure S9 shows TGA (weight loss vs temperature) and derivative weight vs temperature for BM and FD under an argon atmosphere.

Table 4. Shows the weight loss ratio, temperatures, and description of the reaction at each stage in air.

| Sample name | Weight loss % at Temperature 25-500°C | 500- 800°C | Total mass loss % |
|---|---|---|---|
| BM > 300 µm | 8 | 49 | 57 |
| BM 125-300 µm | 8 | 32 | 40 |
| BM < 125 µm | 8 | 25 | 33 |
| FD > 300 µm | 9 | 32 | 41 |
| FD 125-300 µm | 9 | 18 | 27 |
| FD < 125 µm | 12 | 18 | 30 |

Thermal XRD was conducted to investigate changes in solid product phases during heating. A comparison of thermal XRD of BM (Figure 6a) and FD (Figure 6b) > 300 µm fractions in air, increasing temperature to 900 °C, followed by cooling back to room temperature (chosen temperatures: RT, 500, 600, 900, and AFC).

The solid products of the BM were mainly composed of NMC811, Ni, and Mn, and carbon and did not change significantly in the temperature range of 30 to 500°C. Increasing temperature led to a weakening in the diffraction peaks, indicating the decomposition of the cathode material NMC811. When the temperature reaches above 500°C, the XRD pattern of NMC811 disappears, whereas the XRD peaks representing low valent NiO and CoO gradually strengthen. Figures S10 and S11 contain equivalent results for the 125-300 µm (Figure S10a) and < 125 µm (Figure S11a) samples, which show similar trends.

The NMC811 peak observed at 18.5° decreases as temperature increases and disappears above 500°C, suggesting NM811C starts decomposing at temperatures above 500°C. At the same temperature, there is also a decrease in the intensity of the graphite peak (26°) and the secondary NMC811 peaks (43° and 54°). Furthermore, peaks at 37.77° and 38.25° attributed to (LiCo, LiMn oxides, and LiNi) decreased in their intensity, corresponding to the formation of their metals, resulting in the weight loss seen in the TGA, Figure 5. BM < 125 µm has a higher intensity of NMC811 peaks than the BM > 300 µm.

The XRD pattern after 900°C remains almost constant after cooling. The final phases can be summarised as MnO, Co/Ni, $SiO_2$, and Cu. Similarly, [27] investigated thermal treatment of BM derived from LCO and NMC811 cathodes and reported that heating to 1000°C resulted in the formation of various metal oxides, supporting the observed high-temperature transformations.

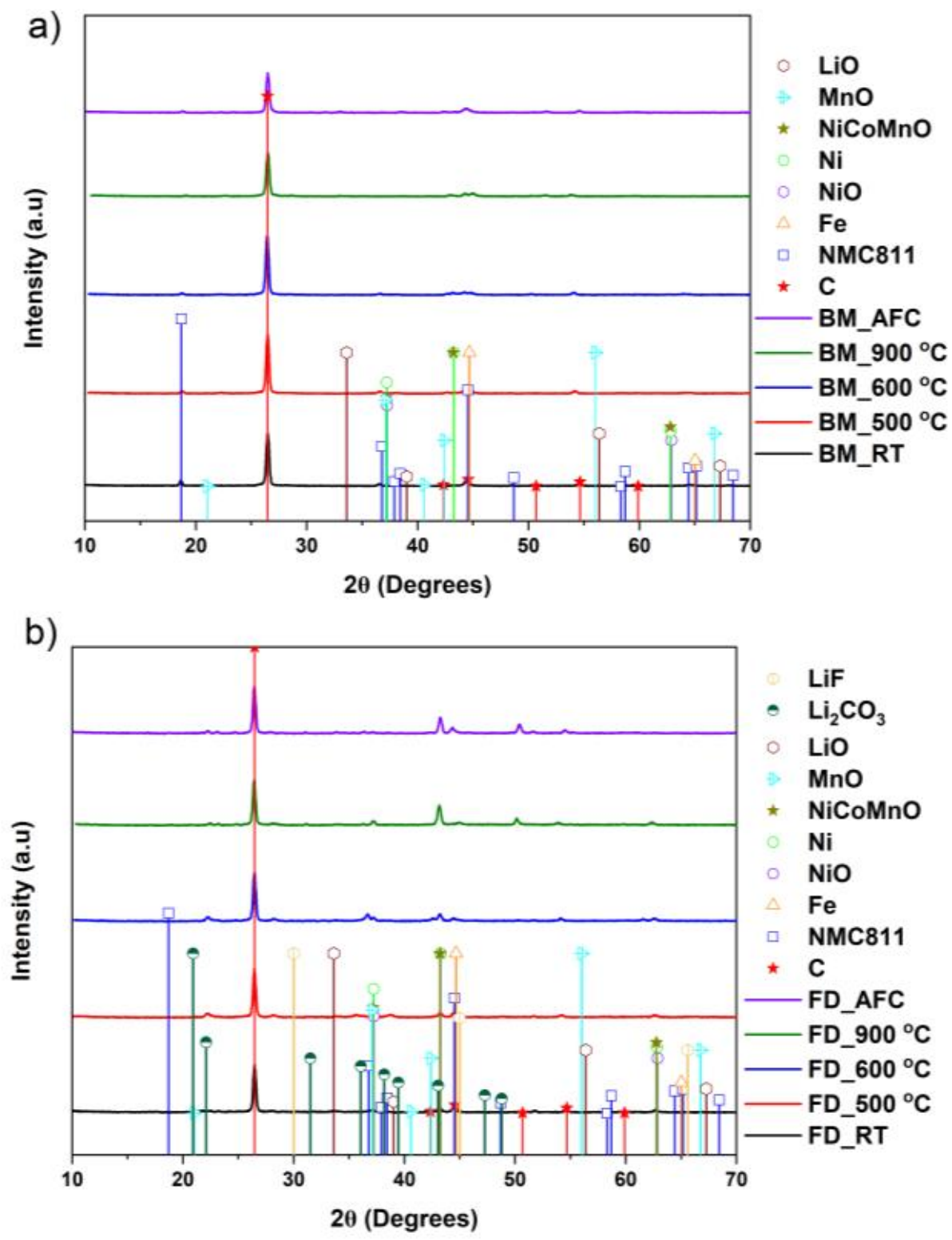

Figure 6. Thermal XRD pattern (RT, 500°C, 600°C, 900°C, and AFC 'after cooling back to room temperature') a. > 300 µm particle size BM, b. FD > 300 µm particle size. *RT and AFC refer to room temperature, and after cooling back to room temperature.

To further understand the environment to which FD material has been exposed, thermal XRD was performed for all fractions using the same conditions: three size fractions (> 300, 125-300, and < 125 µm) heated from 25 to 900°C and cooled to laboratory temperature in an air and $N_2$ atmosphere. Little or no NMC811 is present in Figure 6b; however, FD materials do exhibit metal peaks for Ni and Co. This suggests that the analyzed fire-damaged (FD) material experienced exposure to temperatures exceeding approximately 500°C. In addition to the characteristic reflections of layered NMC811 (R-3m), minor peaks at approximately 23° and 28° (2θ) were observed in some FD subsamples (Figures S12–S14), which can be assigned to $Li_2CO_3$ (110) and LiF (111), respectively. The presence of these phases is consistent with surface reactions between Li-containing oxides and gaseous $CO_2$, as well as $LiPF_6$ decomposition during high-temperature fire exposure.

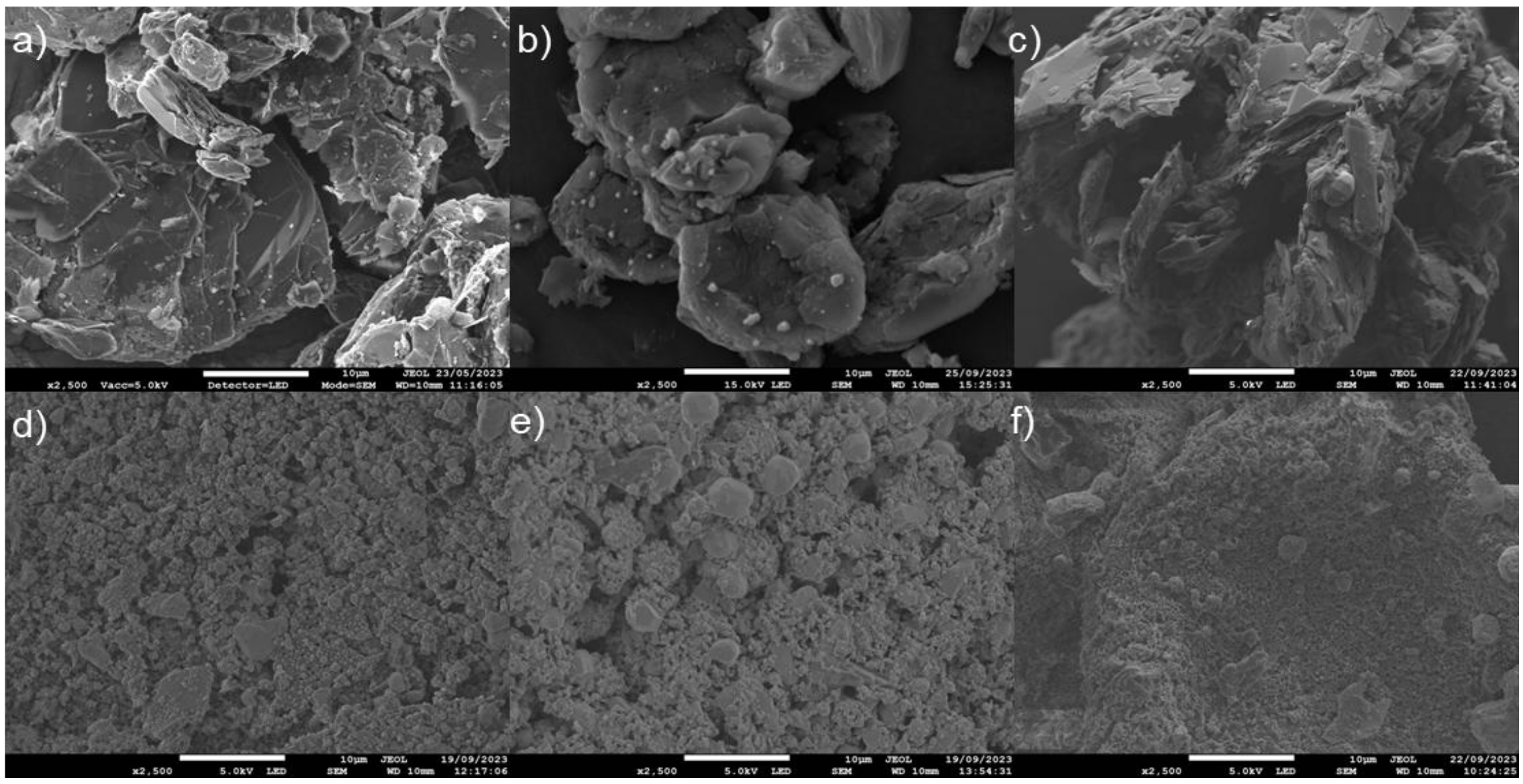

Figure 7: SEM images of BM and FD material after thermal XRD (25°C to 900°C and cooled to laboratory temperature) in an air atmosphere at three fractions (> 300, 125-300, and < 125 µm) (a, b, c) BM, and (d, e, f) FD. This data was acquired by JSM-7800F SEM.

Taken together, the thermal XRD results of the BM and FD materials indicate that the supplied NMC811 fire-damaged feedstock underwent significant thermal alteration, consistent with partial decomposition of the layered oxide at temperatures above ~500°C. Residual graphitic carbon and metal oxide phases remain present, indicating retained material value relevant to recycling. Variations observed among FD subsamples likely reflect heterogeneous thermal exposure within the fire-damaged feedstock, which was obtained without detailed fire metadata. Because the FD material was supplied without detailed information on fire conditions or the number of source cells, these findings should be interpreted as representative of this specific NMC811 fire-damaged feedstock rather than all battery fire scenarios.

Figure 7 shows the SEM of BM and FD material at three fractions (> 300, 125-300, and < 125 µm) after thermal XRD. This step is to better understand the morphology of the FD material after exposure to a high temperature (close to what happens at a fire incident). The NMC811 sphere particle shape disappears (or is difficult to identify) in both samples (BM and FD) at all fractions. Irregularly shaped particles were found in all the samples after thermal XRD, with a size range of sub-micron to around 100 micrometres. SEM of BM and FD samples after thermal XRD at $N_2$ shows the same trend as after thermal XRD at an air atmosphere, Figure S15 [28 - 31].

## Conclusions

Cobalt, manganese and nickel-rich NMC811 particles are predominantly found in BM particles < 125 µm, which can reach 85 % metal content by mass. A similar trend is found in the FD material, but to a lesser extent. Despite indications that the fire, which damaged the batteries, reached > 500°C, evidenced by the breakdown of the NMC811 structure, some organic binders remained in the FD battery materials, suggesting that temperature was not uniform across the battery during the fire.

Fire appears to have caused localized overheating, and a reductive, carbon-rich atmosphere likely promoted partial reduction of $Ni^{3+}$ to $Ni^{2+}$ or metallic Ni. These reduced Ni-rich domains tend to sinter and coalesce into larger, denser agglomerates, resulting in a higher nickel content in the coarse (> 300 µm) fraction of the FD sample. After FD, the batteries are friable and, as such, readily break into small particles without much energy. Whilst this has some benefits in producing small particles, the altered microstructure and formation of Ni-rich aggregates is likely to make the separation of aluminium and copper current collectors from the other materials difficult and therefore increase the quantities of these elements after processing. The incorporation of fire-damaged batteries into BM processing lines is likely to cause significant variability in output product, increasing copper and aluminium whilst reducing the mass percentage of the more valuable NMC811 particles. It is therefore suggested that fire-damaged batteries be processed separately to avoid reducing the value of the standard material. Alternatively, separating the material into different size fractions could enable the chemistries of each fraction to be adjusted 'tuned' to meet different customer specifications. These conclusions apply specifically to NMC811 chemistry, which contains high-value transition metals such as nickel and cobalt, making dilution particularly detrimental. Other chemistries—for example, LFP ($LiFePO_4$)—do not contain Ni or Co and exhibit different thermal degradation behaviours under fire conditions. As a result, the extent of "value loss" and the physical segregation effects observed here may not directly generalize to LFP or other cathode systems. Further work is required to assess how fire damage influences black mass derived from alternative chemistries.

## Acknowledgements

We gratefully acknowledge funding from the EPSRC ECR Fellowship NoREST EP/S03711X/1 (W-AS, JB) and TReFCo EP/W019167/1 (W-AS, RC, DD, JB). The authors would like to thank the access to characterisation equipment to Swansea University Advanced Imaging of Materials (AIM) facility, which was funded in part by the EPSRC (EP/M028267/1) and the European Regional Development Fund through the Welsh Government (80708).